# Gamma-ray spectra of methane in the positron-electron annihilation process

## Xiaoguang Ma[†*] and Feng Wang[*]

*eChemistry Laboratory, Faculty of Life and Social Sciences, Swinburne University of Technology, PO Box 218, Hawthorn, Victoria 3122, Australia*

*E-mail: fwang@swin.edu.au, hsiaoguangma@188.com

**HIGHLIGHTS**

The study reveals that positrophilic molecular electrons of the target molecule, rather than all valence electrons in the molecule, dominate the annihilating process and the Doppler-shift of the gamma-ray spectra of methane in gas phase.

**ABSTRACT**

Bound electron contribution to the Doppler-shift of gamma-ray spectra in the positron-electron annihilation process of molecular methane has been studied in gas phase. Two accurate *ab initio* quantum mechanical schemes, i.e. the delocalized molecular orbital (MO) and the localized natural bond orbital (NBO) schemes, are applied to study the multi-centred methane molecule. The present *ab initio* calculations of methane indicate that the C-H bonds are polarized with the partial negative charge of -0.36 a.u on the carbon atom and the partial positive charge of +0.09 a.u. on each of the hydrogen atoms. The positively charged hydrogen atoms produce repulsive Coulomb potentials to a positron. Both the MO and NBO schemes further reveal that the $2a_1$ electrons of methane, that is, the $2a_1$ electron component of the C-H bonds rather than the whole C-H bonds of methane, predominates the positron-electron annihilation gamma-ray spectra of the molecule. Electrons of a molecule which are dominant the positron-electron annihilation processes are called "positrophilic" electrons in the present study. It is further shown that the negative electrostatic potential (ESP) of methane facilitates with the density of the "positrophilic" $2a_1$ electrons of methane. Other valence electrons (e.g. $1t_2$) in the C-H bonds play a minor "spectator" role in the annihilation process of methane.

---

[†] Permanent address: *School of Physics and Optoelectronic Engineering, Ludong University, Yantai, Shandong 264025, PR China.*

## 1. Introduction

The Doppler-broadened $\gamma$ – ray spectra for positron annihilation in many molecules including hydrocarbons have been measured extensively with milestone achievements (Gribakin et al., 2010) in the past many decades by several experimental groups (Gribakin et al., 2010; Iwata et al., 1997; Tang et al., 1992; Danielson et al., 2012; Kerr et al., 1965). The development of theory in polyatomic molecules has been behind (Crawford, 1994; Jensen and Weiss, 1990; Schrader and Wang, 1976). Recent years, significant theoretical studies have improved our understanding of the $\gamma$ – ray spectra of atoms and small molecules (Dunlop and Gribakin, 2006; Armour and Carr, 1997; Ghosh et al., 1994; Chuang and Hogg, 1967; Wang et al., 2010; Green et al., 2012; Wang (a) et al., 2012; Wang (b) et al., 2012; Iwata et al., 1997). Theoretical development of the annihilation processes for larger polyatomic molecular systems and their chemical effects become paramount important (Danielson et al., 2012).

Molecules exhibit fundamental differences from atoms. First, molecules are multi-centred systems, which result in significantly more difficult to solve the Schrödinger equation quantum mechanically (mathematically), than atomic systems. Almost all the single centred approaches in atomic systems become inappropriate in molecules. Second, molecules form chemical bonds where the valence electrons in molecules are largely delocalised, which do not possess any analogous to atomic systems. Third, polyatomic molecules usually do not have "shells" like atoms but clustered "clouds" on the molecular frame. And finally, certain electronic properties such as dipole moment, partial charges, isomers, resonant structures etc. only apply to molecules.

Historically, quantum mechanics has been developed from atomic systems due to the simplicity. However, it does not imply that all the finding from atomic systems can be directly "mapped" over to molecular systems. For example, the momentum distributions (MDs) of an atomic orbital exhibit as "s-like" if the orbital MDs exhibit a half bell-shape decreasing from the maximum; or as "p-like" if the orbital MDs display a full bell shape. This concept has been applied to molecules for years until the study of diborane ($B_2H_6$) (Wang et al., 2006) when Wang and Pang noted that the half and full bell shaped molecular orbital MDs indicate none and one of nodal plane of the orbitals, respectively, which has little to do with *s* or *p* electrons.

The efforts to reveal the process of annihilation of positron and electron in polyatomic molecules began nearly half a century ago. In 1967, Chuang and Hogg developed a method based on analytic self consistent field (SCF) wavefunctions of carbon and hydrogen atoms to study the momentum distributions of hexane ($C_6H_{14}$) and decane ($C_{10}H_{22}$) in the two photon annihilating positron-electron process (Chuang and Hogg, 1967). It was a significant achievement at the time due to the limited knowledge of quantum mechanics and computer resources for larger molecules. It was concluded (Chuang and Hogg, 1967) that positrons annihilate almost exclusively with electrons in the C-H and C-C bonds of hexane and decane.

Applications of the analytic SCF wavefunction method (Chuang and Hogg, 1967) to other polyatomic molecules have experienced substantial restrictions as it is a molecular specific rather than a robust method. The first obstacle is that the analytic wavefunctions of the target molecule need to be derived for the particular molecule under study each time. As a result, obtaining accuracy and robust molecular wavefunctions of polyatomic molecules have been a bottleneck to study gamma-ray spectra of molecules. The second obstacle is that larger hydrocarbons such as hexane and decane contain both C-H and C-C bonds; it is not clear which group of bonds, the C-H or C-C bonds, dominates the gamma-ray spectra, or whether C-H and C-C bonds contribute equally to the gamma-ray spectra of hydrocarbons.

The first bottleneck of accurate and robust target molecular wavefunctions is partly resolved with the application of *ab initio* computational chemistry methods (Wang et al., 2010; Green et al., 2012; Wang (a) et al., 2012; Wang (b) et al., 2012; Iwata et al., 1997). Previous studies suggest that dominant contribution to the profiles of the $\gamma$-ray spectra is from certain electrons rather than all valence electrons (Wang et al., 2010; Green et al., 2012; Wang (a) et al., 2012; Wang (b) et al., 2012; Iwata et al., 1997). The present study targets the second bottleneck, that is, the C-H and/or C-C bonds. Methane ($CH_4$) is the smallest alkane which possesses only the C-H bonds. It has been the prototype molecule in the history of chemical research (Wang 2004; Gray and Robiette, 1979). Such structural characteristics of methane enable us to study the C-H bonds in the absence of the C-C bonds.

## 2. Methods and computational details

The wavefunctions of the electrons in different orbitals or bonds of methane have been calculated using the Gaussian09 computational chemistry package (Frisch et al., 2009). In the *ab initio* Hartree-Fock calculations, the TZVP basis set (Schaefer et al., 1994) is used. In this basis set the atomic carbon orbitals are constructed by the C(5s9p6d) scheme of Gaussian type functions (GTFs), while the atomic hydrogen orbitals are constructed by the H(3s3p) scheme of GTFs. The methane molecular wavefunctions are calculated using the HF/TZVP model (Schaefer et al., 1994) quantum mechanically. The structural parameters (bond lengths and bond angles) of methane ($CH_4$) are based on literature values (Wang 2004; Gray and Robiette, 1979).

In order to identify which valence electrons of methane dominantly annihilate with the positron, the wavefunctions of each electron in methane are obtained using two quantum mechanical schemes. One is the conventional molecular orbital (MO) scheme and the other is the natural bond orbital (NBO) scheme (Weinhold and Carpenter, 1988). The former presents a delocalized valence electron picture, whereas the latter localizes the valence electrons onto the C-H bonds of methane. In the MO scheme, the ground state electronic configuration of methane contains five doubly occupied molecular orbitals as $(1a_1)^2(2a_1)^2(1t_2)^6$, which consists of two $1a_1$ core electrons and eight valence electrons of two $2a_1$ and six $1t_2$ (three-fold degenerate) electrons. In the alternative NBO scheme, the electronic configuration of methane is partitioned into two core electrons and eight localized valence electrons in four equivalent C-H bonds of methane.

If an annihilation occurs, the probability of annihilating an electron from orbital $i$ of a molecule can be estimated by the molecular orbital wavefunction (Crawford, 1994). Positrons are unlikely to annihilate core electrons of atoms and/or molecules. As a result, the present study concentrates on the eight valence electrons of methane, that is, the electrons constitute the four equivalent C-H bonds in the NBO scheme, or alternatively, the two $2a_1$ and six $1t_2$ electrons in the MO scheme. The calculated wavefunction $\psi_{CH}(r)$ of the C-H electrons in methane is dominated by the carbon 2s and 2p electrons as well as the hydrogen 1s electrons:

$$\psi_{CH}(r) \equiv \psi_C(r-R_C)[\approx 2s(14.28\%) + 2p(44.80\%)] \\ +\psi_H(r-R_H)[\approx 1s(40.80\%)], \qquad (1)$$

where $\psi_C(r-R_C)$ and $\psi_H(r-R_H)$ are the wavefunctions centred on the carbon atom and the hydrogen atoms, respectively, from the NBO scheme. The percentage weights of the atomic carbon orbitals are obtained from the output of the *ab initio* calculations. Alternatively, in the MO theory, the valence electron wavefunctions are given by,

$$\psi_{2a_1}(r) \equiv \psi_C(r-R_C)[\approx 2s(82\%)] \\ + \psi_H(r-R_H)[\approx 1s(14\%) + 2p(4\%)], \quad (2)$$

and

$$\psi_{1t_2}(r) \equiv \psi_C(r-R_C)[\approx 2p(67\%)] \\ + \psi_H(r-R_H)[\approx 1s(33\%)]. \quad (3)$$

Where the $1t_2$ wavefunction is three fold energy degenerated and the highest occupied molecular orbital (HOMO). The C-H wavefunctions in Eq.(1) represent the valence electrons of methane in the NBO scheme, which must be equivalent to the valence electrons ($2a_1$ electrons in Eq.(2) and $1t_2$ electrons in Eq.(3)) in the MO scheme, as both schemes represent the same valence electrons of methane. The valence electron wavefunctions of methane are directly mapped into the momentum space (Ferrell, 1956), from which the spherically averaged $\gamma$-ray spectra for each type of electrons are calculated (Dunlop and Gribakin, 2006; Chuang and Hogg, 1967).

## 3. Results and discussion

It is known that Coulomb potential and electron density play important roles in positron annihilation (Gribakin et al., 2010; Iwata et al., 1997; Tang et al., 1992; Danielson et al., 2012). A positron annihilates one electron each time, the probability of annihilating an electron is determined from the molecular orbital (MO) where the electron resides (Crawford, 1994). If a sufficient number of electrons in polyatomic molecules, then the bound electrons will dominate the annihilation. Recently, Surke *et. al.* suggested that the attractive potential in molecules increases probability of annihilation (Danielson et al., 2012). Molecules without permanent dipole moment such as alkanes do not guarantee that the electron density is the same over the entire molecule. For example, it is a known fact that the proton affinity of the centre C-C bond in straight-chain n-alkanes, such as n-hexane is the highest (Hunter and East, 2002). In the case of methane, although it does not have a permanent dipole moment, all the C-H bonds are polar bonds. As a result, it is critical for an accurate and detailed *ab initio* quantum mechanical study of methane, in order to understand the role of the electrons in the C-H bonds of methane in the positron-electron annihilation process.

Fig.1 compares the calculated individual valence electron contributions, based on the two different schemes, to the $\gamma$-ray spectra with the two-Gaussian fitted experiment (Iwata et al., 1997). Each orbital or bond exhibits its unique characteristic linewidth and profile to the $\gamma$-ray spectra of methane, which has been observed in our previous studies (Wang et al., 2010; Wang et al., 2012). It is clear that the $\gamma$-ray profile of the $2a_1$ electrons (black solid line) best fits the experimental profile of methane. This is particularly the case in the low energy region of photon energy shift is smaller than 3 keV. Fig.1 also shows that frontier valence electrons (i.e. the $1t_2$ electrons) on the HOMO (red dash line) exhibit apparently different profiles from the measurements, indicating that the HOMO valence electrons of methane do not dominate the contributions, in agreement with Crawford (Crawford, 1994) that calculations for hydrocarbons suggest "most of the time the hole is created in an MO below the highest occupied molecular orbital (HOMO)." In the case of methane, the "hole" orbital can only be the $2a_1$ orbital.

The total valence electrons of methane, either in the C-H bonds (cyan dash dot line) or in the total valence ($2a_1+1t_2$) electrons (blue dot line), show large discrepancies to the experiment. This suggests that the probability of annihilation of a positron is not the same to all valence electrons in a molecule. Similar observations are obtained from positron-annihilation-induced Auger-electron spectroscopy that positrons can become localized at defects of the surface, which is different from average of the whole surface (Jensen and Weiss, 1990). As in such broad ensembles the spectroscopic information of an individual orbital is washed out or manipulated by all valence electrons. Hence, important contributions of the dominant orbital(s) are buried under the inhomogeneous broadening. Finally, Fig.1 also clearly exhibits the equivalence between the NBO and MO schemes---the C-H profile with a full width at half maximum (FWHM) of 2.85 keV and the valence electrons in the MO scheme with the same FWHM.

It has been well understood for many years that the Coulomb repulsion of the nuclei keeps the positron from approaching close to the target molecule. Although methane has high symmetry without permanent dipole moment, the C-H bonds are polar bonds with a partial negative charge of –0.36 a.u. on the central carbon atom and partial positive charge of +0.09 a.u. on each of the hydrogen atoms. As a result, the terminal hydrogens of methane are partially naked nuclei, which are unlikely to attract the positively charged positron.

Coulomb potential controls interactions between charged particles in the vicinity. To further reveal the reason why $2a_1$ electrons dominate the annihilation, the total molecular electrostatic potential (ESP) of methane is calculated *ab initio*. The ESP serves as an indicator for the nucleophilic and electrophilic forces of the molecule in three-dimensional space (Murray et al., 1994; Politzer and Murray, 2002). Fig.2 reports the calculated ESP of methane. The positive area (red dot) of the ESP implies the dominance of nuclei which is electrophilic but repulsive for positive charges. The negative area (blue solid) of the ESP implies the dominance of electrons which is attractive for positive charges, i.e. nucleophilic or "positrophilic" in present study. Four equivalent negative potential cone regions locate on the ESP, which do not locate on any atoms, neither the carbon nor the hydrogens. This represents the calculated lowest attractive potential trajectory for a positron to approach the carbon in methane.

It is proposed that in the annihilation process, the probability of positron to approach a target in any directions in gas phase is the same when it is far away from the target, but the trajectory of the positron will change to follow the ESP of the target when the positron is sufficiently close to the target. Fig.3 provides a schematic illustration that a positron approaches the target in the direction of (maximum) positive (Fig.3a) and negative (Fig.3b) ESP of methane as indicated by the orange arrow. The calculated valence electron densities of the $2a_1$, $1t_2$ and the C-H electrons are indicated in the figures, together with the repulsive (red dot line, Fig.3a) and attractive (blue dot line, Fig.3b) potential energy curve along the trajectory. In the figures, the vertical axis at 1.65 Å is where the zero ESP position (see Fig.3(b)). As seen in Fig.3a, the valence electrons are screened by the total positive ESP (red dot curve) of methane, regardless the electron densities presented in this region. That is, the partially positively charged hydrogen of methane is the dominant repulsive force in the position. The positron in Fig. 3a will be bounced back when it hits the ESP repulsive wall (blue dot curve) in this direction so that it unlikely to annihilation an electron of the target.

A positron is accelerated to the target due to the Coulomb attractive force field (an attractive potential well of ESP) in Fig.3b, however. In this scenario, the positron is attracted by the Coulomb attractive as shown in the potential energy curve (blue) and is accelerated. It slows down when the potential energy increases as shown in the repulsive curve (blue dot curve) in Fig.3b. The electron density also increases quickly so that the positron continues to approach the target until it annihilates the electron. For methane, the positron most likely annihilates the $2a_1$ valence electrons, which happens to be the first valence electrons as shown in Fig.3b, the "positrophilic" electrons. In $\gamma$ – ray spectra of methane, the positrophilic electrons are the $2a_1$ electrons. The finding agrees with (Crawford, 1994).

## 4. Concluding remarks

In summary, the results from the competition between the ESP and the density of a valence electron contribute to $\gamma$ – ray spectra of the methane. The present study demonstrates that the dominant contribution is from the positrophilic valence $2a_1$ electrons or the $2a_1$ component of the C-H bonds in methane, rather than other valance electrons of methane the first time. It is in agreement with an early finding of hydrocarbons that "the hole is created in an MO below the highest occupied molecular orbital (HOMO)" of (Crawford, 1994). Hydrogen atoms which are partially positively charged in the C-H bonds of methane are unlikely the positrophilic sites in annihilation. The present study further rationalizes that the positrophilic electrons of methane are the $2a_1$ valence electrons using ab initio calculations of Coulomb electrostatic potential of methane. More evidences on other molecules, such as n-hexane will be published elsewhere.

## Acknowledgements

This project is supported by the Australia Research Council (ARC) under the discovery project (DP) scheme. National Computational Infrastructure (NCI) at the Australia National University (ANU) under the Merit Allocation Scheme (MAS) is acknowledged.

## References

Armour, E. A. G., Carr, J. M., 1997. *J. Phys. B* 30, 1611.

Chuang, S. Y., Hogg, B. G., 1967. *Can. J. Phys.* 45, 3895.

Crawford, O. H., 1994. *Phys. Rev. A* 49, R3147.

Danielson, J. R., et al., 2012. Phys. Rev. Lett. 109, 113201.

Dunlop, L. J. M., Gribakin, G. F., 2006. *J. Phys. B* 39, 1647.

Ferrell, R. A., 1956. *Rev. Mod. Phys.* 28, 308.

Frisch M J et al 2009. Gaussian 09, Gaussian, Inc., Wallingford, CT

Ghosh, A. S., et al., 1994. *Hyperfine Interactions* 89, 319.

Gribakin, G. F., et al., 2010. *Rev. Mod. Phys.* 82, 2557.

Green, D. G., et al., 2012. *New J. Phys.* 14, 035021.

Gray, D. L., Robiette, A. G., 1979. *Mol. Phys.* 37, 1901.

Hunter, K. C., East, A. L. L., 2002. J. Phys. Chem. A 106, 1346.

Iwata, K., et al.,1997. *Phys. Rev. Lett.* 79, 39.

Iwata, K., et al., 1997. *Phys. Rev. A* 55, 3586.

Jensen, K. O., Weiss, A., 1990. *Phys. Rev. B* 41, 3928.

Kerr, D P., et al., 1965. *Mol. Phys.* 10, 13.

Murray, J. S., et al., 1994. *J. Mol. Struct. (THEOCHEM)* 307, 55.

Politzer, P., Murray, J. S., 2002. *Theor. Chem. Acc.* 108, 134.

Schrader, D. M., Wang C. M., 1976. *J. Phys. Chem.* 80, 2507.

Schaefer, A., et al., 1994. *J. Chem. Phys.* 100, 5825.

Tang, S., et al., 1992. *Phys. Rev. Lett.* 68, 3793.

Wang, F., et al., 2010. *J. Phys. B* 43, 165207.

Wang(a), F., et al., 2012. *E. J. Phys. D* 66, 107.

Wang(b), F., et al.,2012. *New J. Phys.* 14, 085022.

Wang, F., et al., 2006. *J. electron. Spectrosc. Relat. Phenom.* 151, 215.

Wang, F., 2004. *J. Mol. Stru.(Theochem)* 678, 105.

Weinhold, F., Carpenter, J. E., 1988. *The Structure of Small Molecules and Ions* (R. Naaman, and Z. Vager, Plenum)

Figure captions

Figure 1: Gamma-ray spectra of methane molecule in positron-electron annihilation process compared with two-Gaussian fitted experimental data (○). The solid line represents the $2a_1$ electrons, the red dash line represents the $1t_2$ electrons; the blue dot line represents the total valence electrons, and the cyan dash dot line represents the electrons in C-H bonds. All spectra are normalized to unity at zero. The numbers in brackets are the Full Widths at Half Maximum (FWHM) in keV unit.

Figure 2: The total electrostatic potential (ESP) in three dimension space. The red dot surface has a positive potential which repulses a positron while blue solid surface has a negative potential which attracts a positron. The net charge at each atom has been given.

Figure 3: (a) The scheme that a positron is unlikely to annihilate with electrons. The values of the ESP and electron density curves are taken along the direction of the proposed positron attacking. (b) The scheme that a positron is likely to annihilate an electron. The values of the ESP and electron density curves are taken along the direction of the positron approach. The olive short dot horizontal line represents the zero potential.

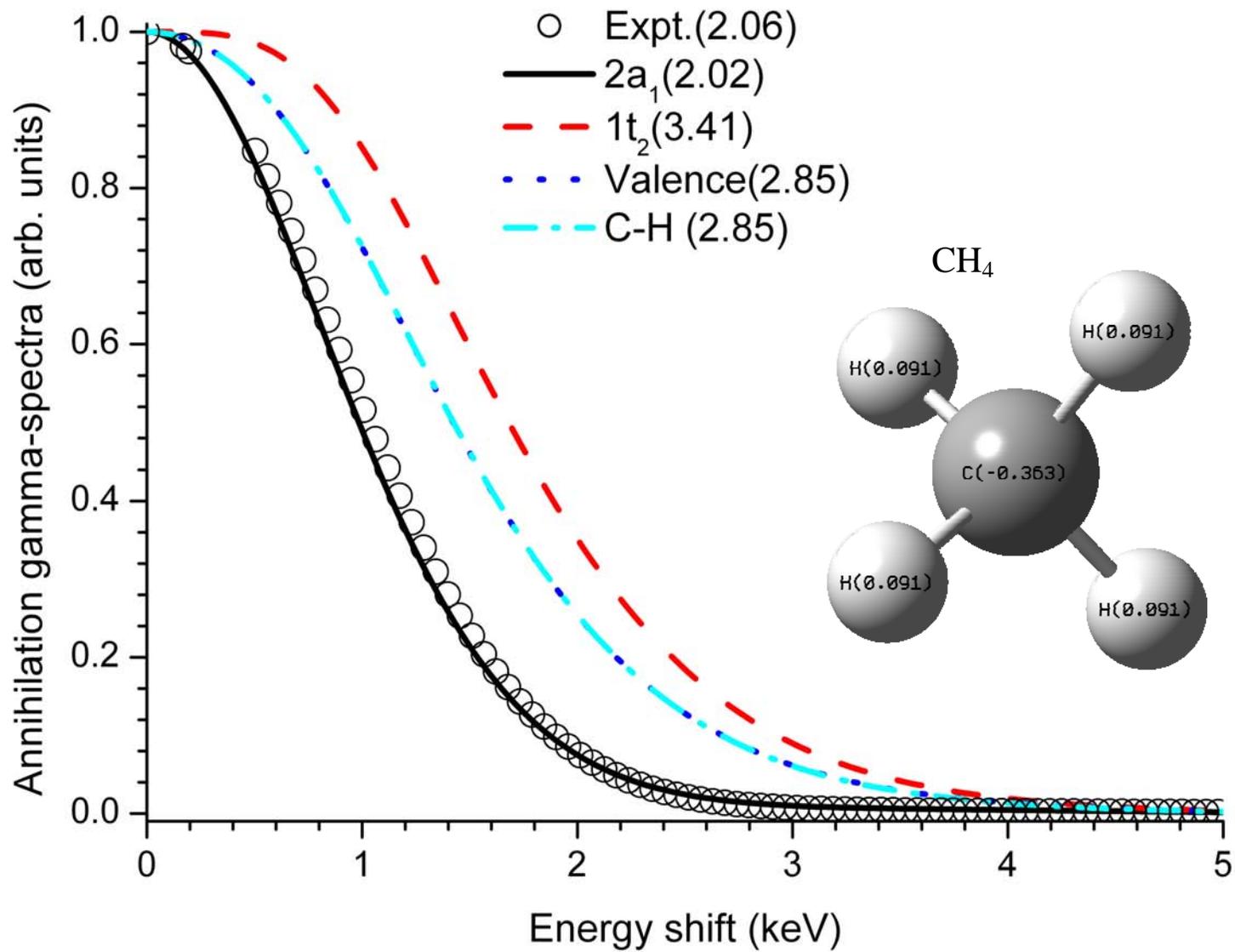

Fig.1

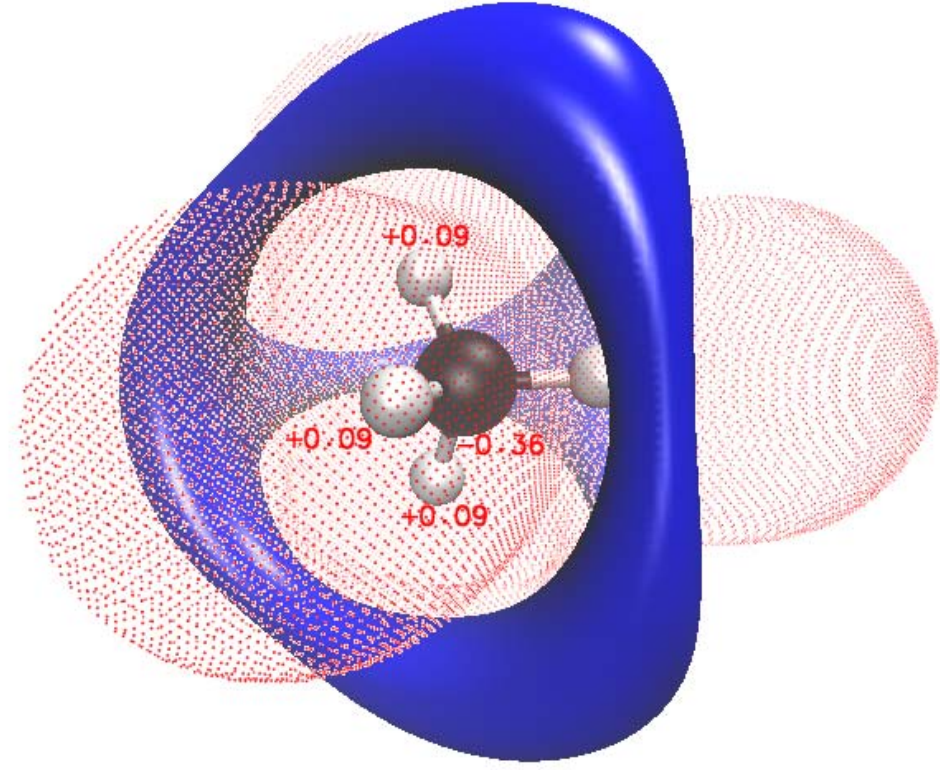

Fig.2

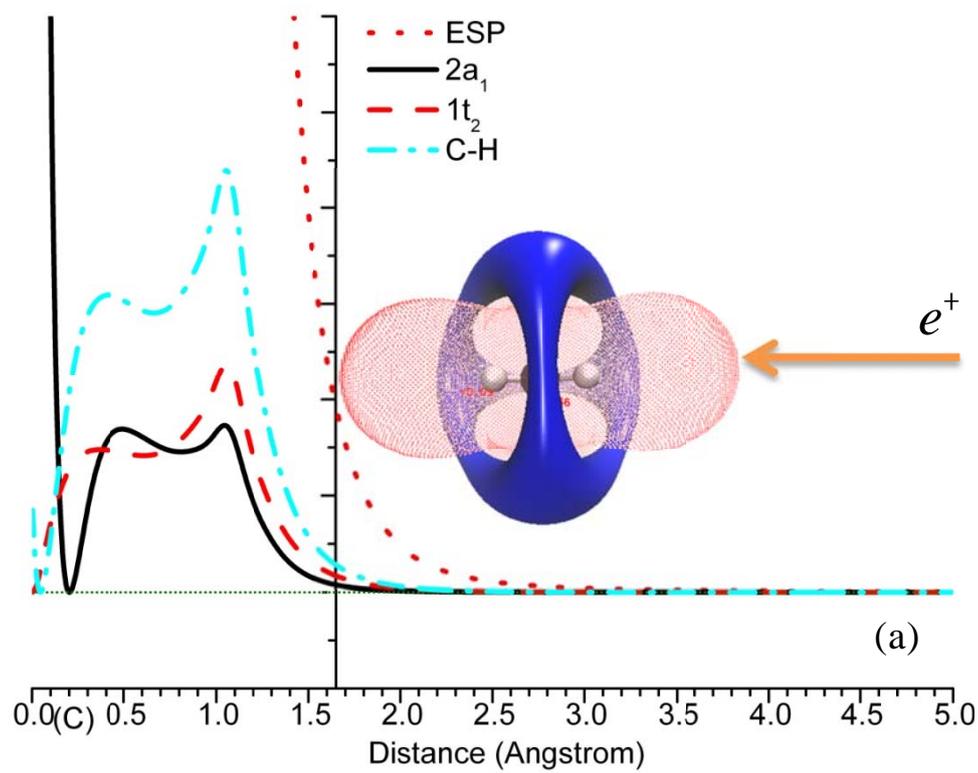

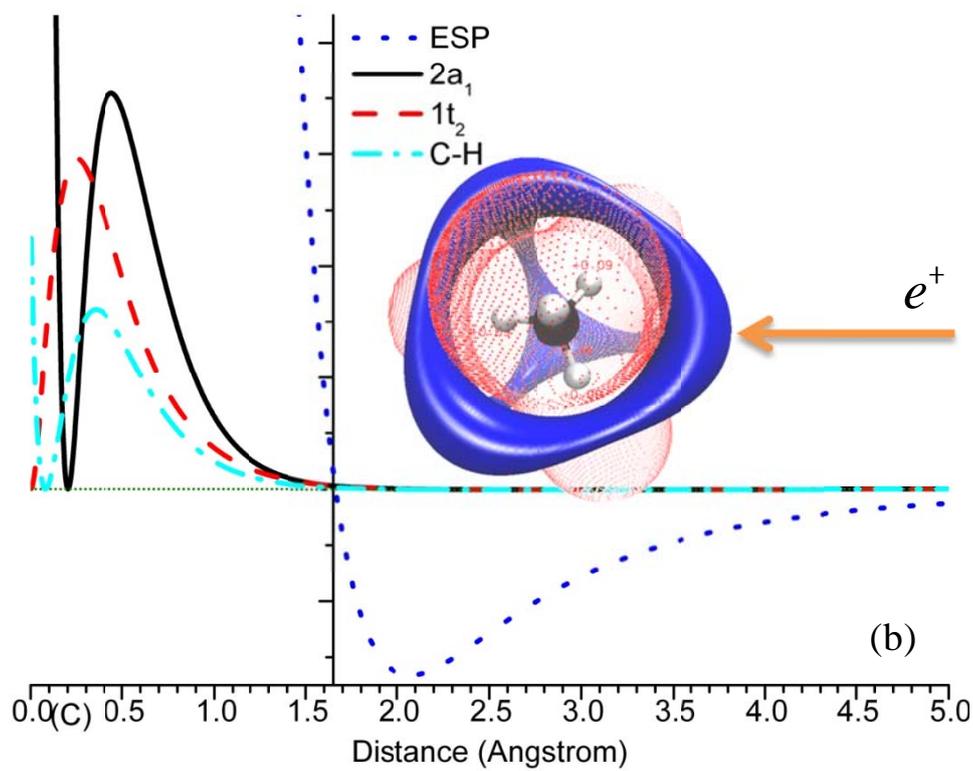

Fig.3